\documentclass[prl,twocolumn,showpacs]{revtex4}

\usepackage{graphicx}

\begin{document}

\newcommand{\etal}{{\it et al.}\/}
\newcommand{\gtwid}{\mathrel{\raise.3ex\hbox{$>$\kern-.75em\lower1ex 
\hbox{$\sim$}}}}
\newcommand{\ltwid}{\mathrel{\raise.3ex\hbox{$<$\kern-.75em\lower1ex 
\hbox{$\sim$}}}}

\title{Pairing on striped $t$-$t'$-$J$ lattices}

\author{Steven R.~White}
\affiliation{Department of Physics and Astronomy, University of California,
Irvine, CA 92697-4575 USA}
\author{D.~J.~Scalapino}
\affiliation{Department of Physics, University of California,
Santa Barbara, CA 93106-9530 USA}

\date{\today}

\begin{abstract}

Results are given from a density matrix renormalization group
study of pairing on a striped $t$-$t'$-$J$ lattice in the
presence of  boundary magnetic and pair fields. We find that
pairing on a stripe depends sensitively on both $J/t$ and $t'/t$.
In the strong-pairing model-parameter regime the 
stripes are easily coupled by the pair field, and have a uniform phase.
There is a small but measurable energy cost to
create anti-phase  superconducting domain walls. 

\end{abstract}

\maketitle

Experimental studies are providing new insight into the interplay
of the charge, spin and $d$-wave pairing correlations in the
underdoped cuprates.  Scanning tunneling microscopy
measurements\cite{ref:Kohsaka} on Ca$_{1.88}$Na
$_{0.12}$CuO$_2$Cl$_2$ and
Bi$_2$Sr$_2$Dy$_{0.2}$Ca$_{0.8}$Cu$_2$O$_{8+8}$, suggest that  at
low temperatures, as the doping increases, superconducting
correlations  develop on a glassy array of 4a$_0$ wide domains
oriented along the Cu-O $x$ or  $y$-bond directions.
Neutron\cite{ref:Tranqueda} and x-ray scattering\cite
{ref:Abbamonte} experiments on La$_{1.873}$Ba$_{0.125}$CuO$_4$
find charged stripes with a 4a$_0$  period separated by
$\pi$-phase shifted antiferromagnetic regions.  Moreover, when
the temperature decreases below the spin ordering temperature,
the  planar $\rho_{ab}$ resistivity shows evidence of a
Kosterlitz-Thouless\cite {ref:Kosterlitz} like behavior
consistent with the development of 2D pairing correlations.\cite
{ref:Li} Remarkably, $\rho_{ab}$ follows the Halperin-Nelson\cite
{ref:Halperin} 2D prediction over an extended temperature range,
implying a decoupling of the  pair phase between the CuO$_2$
planes.  In underdoped La$_{2-x}$Sr$_x$CuO$_4$, superconductivity
and static spin density waves coexist\cite {ref:Lake}, and recent
far-infrared measurements\cite{ref:Schafgans} find that the
Josephson plasma resonance is quenched by a modest magnetic field
applied  parallel to the c-axis. An applied c-axis magnetic field
is known to stabilize a  magnetically ordered
state\cite{ref:Lake} for a range of dopings near $x=1/8$. This is
believed to be a striped state and Schafgans \etal
\cite{ref:Schafgans} have argued that just as in
La$_{1.873}$Ba$_{0.125}$CuO$_4$, the establishment of
antiferromagnetic stripes leads to a suppression of the
interlayer Josephson coupling. To explain this suppression, it
has been suggested that anti-phase domain walls in the $d$-wave
order  paramater, locked to the SDW stripes, are stablized in the
striped magnetic state.\cite{ref:Berg,ref:Himeda} In this case,
the 90$^\circ$ rotation of the stripe order between adjacent
planes would  lead to a cancellation of the interlayer Josephson
coupling.  

Early mean field calculations\cite{ref:Poilblanc} found striped
states in $t$-$J$ and Hubbard models. However, these stripes had
a hole filling which  was twice that which was observed in the
cuprates. While arguments\cite {ref:Machida} were made that this
problem could be overcome by including a next near  neighbor
hopping $t'$, an alternative view suggested that it reflected the
importance of underlying $d_{x^2-y^2}$ pair field
correlations.\cite {ref:WhiteScalapinoArxiv9610104} Density
matrix renormalization group (DMRG) calculations 
found half-filled hole
stripes, $\pi$-phase shifted  antiferromagnetism and short ranged
$d_{x^2-y^2}$ pairing correlations.\cite{ref:WhiteScalapino3} 
This interplay of oscillating
hole density,  spin density and $d_{x^2-y^2}$ superconductivity
was also found in Gutzwiller- projected variational Monte Carlo
(VMC) calculations.\cite{ref:Himeda} 
In certain parameter ranges, these VMC calculations found 
a stable striped  state in which the
$d_{x^2-y^2}$ pair field had $\pi$ phase shifts between the
stripes, i.e. anti-phase domain walls.  A recent
renormalized mean field theory (RMFT) treatment found a similar
small energy difference, with the uniform phase $d$-wave state
lying slightly lower in energy.  \cite{ref:Yang}
A $\pi$- phase shifted $d$-wave pair field would provide a natural
explanation for the observed suppression of the interlayer
Josephson coupling.\cite {ref:Berg,ref:Himeda}
A similar, low lying, modulated
superconducting  state was also found in VMC  resonating valence bond (RVB)
calculations.\cite{ref:Raczkowski,ref:Capello} Here,  however, no
incommensurate AF order was assumed.\cite{ref:foot1} 

We have  
carried out a
series of DMRG calculations on under-doped $t$-$t'$-$J$ lattices  
with a Hamiltonian
\begin{eqnarray}
H=-t\sum_{\langle ij\rangle}(c^\dagger_{is}c_{js}&+h.c.)
      -t'\sum_{\langle ij\rangle'}(c^\dagger_{is}c_{js}+h.c.)\nonumber\\
	    &+J\sum_{\langle ij\rangle}\left(\vec{S}_i\cdot \vec{S}_j-\frac{n_in_j}{4} 
\right).
\label{eq:1}
\end{eqnarray}
Doubly occupied sites are excluded from the Hilbert space,
$\vec{S}_i $ and $c^\dagger_{i,s}$ are electron spin and creation operators,
respectively,  and
$n_i=c^\dagger_{i\uparrow}c_{i\uparrow}+c^\dagger_{i\downarrow}c_{i\downarrow}$
is the electron number on site $i$. There is a near neighbor
$\langle ij \rangle$ hopping $t$, a next near neighbor $\langle
ij\rangle'$ hopping  $t'$, and an exchange coupling $J$. We set $t=1$.
 Using boundary magnetic
and pair fields, we have explored how a pair field is established
over a magnetically striped array at low doping.  

\begin{figure}[htbp]
\begin{center}
\includegraphics[width=6cm,clip,angle=0]{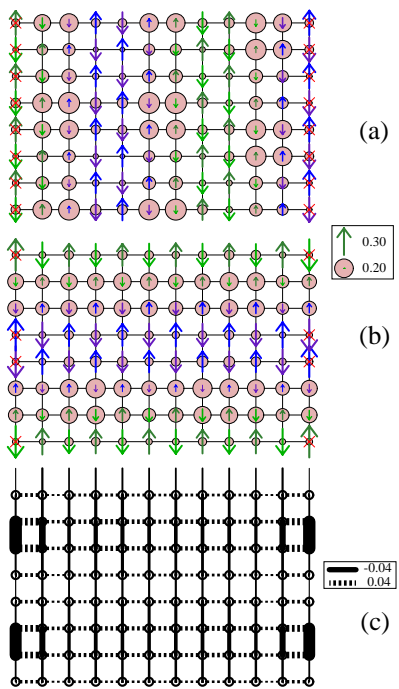}
\caption{(a) Hole $\langle 1-n_i\rangle$ and 
spin densities $\langle S^z_i\rangle$ for a $12\times8$ lattice with 12 holes and
$J=0.5$ and $t'=0$, with  cylindrical  boundary
conditions: periodic in the  $y$-direction, open  in the
$x$-direction. A staggered magnetic field of magnitude $h=\pm0.05$
has  been applied to the ends (red X's). (b) A similar lattice with
$J_x=0.55$, $J_y=0.45$ and  $t'=0.0$ such that the stripes run
along the $x$-axis. Here we use $h=\pm0.2$ 
on the sites with a red X,   and a pair field
$\Delta_0=1.0$ to the edge links without X's.  A
chemical potential $\mu=1.23$ was used to give a doping of
$x=0.127$.
(c) The pair field strength $\langle D_{ij}\rangle $ on each  link for the
system shown in (b).
}
\label{fig:1}
\end{center}
\end{figure}

 Fig.~\ref{fig:1} shows two $12\times8$ ladders (tubes) with cylindrical boundary
conditions (CBCs).
In Fig.~\ref{fig:1}(a), a weak staggered magnetic field is applied
to the open ends and the number of holes is fixed at 12,  
corresponding to a doping
$\delta=0.125$. The end boundary conditions break the translational  
 and spin symmetries in the x-direction, giving rise to
finite, varying values of $\langle n_i\rangle$ and $\langle S^Z_i\rangle$, but 
the basic stripe pattern, with antiferromagnetic
order between stripes and with $\pi$ phase shifts in the magnetic order
across stripes, is  
intrinsic and the
open ends and boundary $h$ field only act to pin the stripes.  Up to $m=4000$
states per block were kept.
 
In this cluster there are four holes per stripe, and DMRG calculations on longer stripes have shown  
that this linear stripe filling of 0.5 holes per unit length is the
preferred filling.\cite{ref:WhiteScalapino3} 
Fluctuations in which a pair of holes are  
exchanged between
the stripes are energetically unfavorable for this short a stripe,
and the local $d_{x^2-y^2}$ pair field correlations are  
short ranged, and pair field coupling between stripes is negligible.

Previous DMRG calculations have not found ground states with
both extended pairing correlations and stripes. There are two reasons:  first, it has been
difficult to construct limited-size clusters allowing significant particle number
fluctuations on a stripe, and second, as we show below,
the model parameters which
strongly favor pairing (e.g. $J/t\sim 0.5$, $t'/t\sim 0.2$) are  different from the values usually
taken to represent the cuprates (e.g. $J/t\sim 0.3$, $t'/t=-0.2$).  In Fig.~\ref{fig:1}(b), we show
results for a cluster which does have both stripes and pairing.
In order to allow hole fluctuations, 
a slightly anisotropic exchange interaction ($J_x=0.55$, $J_y=0.45$) was chosen to  
favor orienting the stripes along the x-direction, overcoming an opposite tendency due to
the cylindrical geometry.
Then, in addition to the magnetic fields at the open left and right ends, a pair field  
coupling has been applied to the ends of the stripes. Defining the link pair creation operator
\begin{equation}
\Delta_{ij}^\dagger=\frac{1}{\sqrt{2}}(
	c^\dagger_{i\uparrow}c^\dagger_{j\downarrow}+c^\dagger_{j\uparrow}c^\dagger_{i\downarrow})
\label{eq:2a}
\end{equation}
on specified links, we add to the Hamiltonian boundary region terms of the form $\Delta_0 D_{ij}$, where
\begin{equation}
  D_{ij}=\frac{1}{2}\left[ \Delta_{ij}^\dagger + \Delta_{ij}\right],
\label{eq:2}
\end{equation}
and we measure $D_{ij}$ on each link in the resulting (approximate) ground state.
For this system we took $\Delta_0=1.0$ for the four thickest lines of (c), and also $\Delta_0=0.5$ for
the four vertical links adjacent to them.
With the pair field boundary
conditions, total particle number is only conserved modulo two, and the average number of holes is 
controlled by a chemical  
potential $\mu$.  Up to $m=6000$ states were kept per block during 23 sweeps in this calculation.
As shown in  
Fig.~\ref{fig:1}(c),
a proximity $d$-wave pair field is established throughout the lattice.

In order to understand the system in more detail, it is useful to separate questions dealing with (a) pairing on
a stripe from  (b) pairing between stripes.  First, can a single stripe support strong pairing, and if so, 
for what model parameters? Second,
do stripes with pairing couple their pair fields, and if so, is the coupling in phase or antiphase? To answer these
questions we will study clusters somewhat smaller than shown in Fig. 1, to ease the computational burden
and increase the accuracy.  

\begin{figure}[thbp]
\begin{center}
\includegraphics[width=8cm,clip,angle=0]{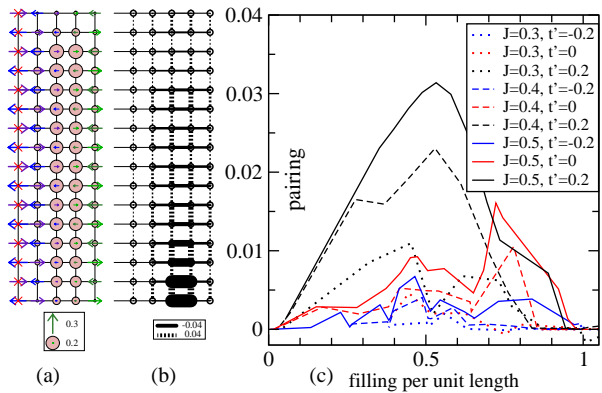}
\caption{ A cluster with a single forced stripe on
a $16\times5$ ladder with CBCs, with
the $x$ direction oriented vertically.  
For all dopings, on the $y=5$ leg, a staggered
field $(-1)^x h$ with $h=0.05$ was applied, along with a chemical
potential of 2.0.  On the other 4 legs, a chemical potential
$\mu$ was applied to vary the doping; in the case shown in (a)
and (b), $\mu=1.41$, $J=0.5$, and $t'=0.2$, yielding a doping of
$x=0.106$, which corresponds to a linear doping of $0.53$.  In
each case a strong pair field was applied to four  rungs ($x=1-4$, connecting $y=2$ and $y=3$), 
visible as the four thickest links in (b).  The
magnitude of the applied field in the four rungs was 1.0, 1.0,
0.5, 0.25.  (a) The hole density $\langle 1-n_i\rangle$ and spin
density $\langle S^z_i\rangle$.  (b) The measured pair field
$\langle D_{ij}\rangle $.  (c) The
measured pair field at the $x=12$, $y=2-3$ rung versus the
number of holes per unit length. The maximum value for $\langle D_{ij}\rangle $ in (c) is of order half of what one 
would find for a BCS ground state with $\Delta_0/t=0.1$.}
\label{fig:2}
\end{center}
\end{figure}

\begin{figure}[thbp]
\begin{center}
\includegraphics[width=4cm,clip,angle=0]{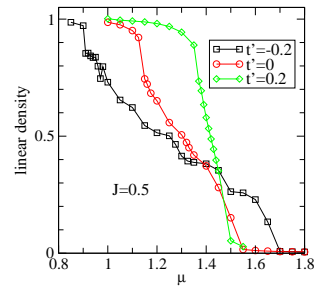}
\vskip 0cm
\caption{The linear density versus $\mu$ for different values of
the next nearest neighbor hopping $t'$ for $J=0.5$ for the
systems of Fig. 2.}
\label{fig:3}
\end{center}
\end{figure}

To look in more detail at the pairing correlations associated
with  a stripe, we have studied a single stripe on the
$16\times5$ lattice with CBCs, shown in  Fig.~\ref{fig:2}.  In this
case boundary conditions were used to force the presence of
a stripe similar to those shown in
Fig.~\ref {fig:1}(b)-(c).  Here a strong pair field was applied to only
one end of the stripe. The proximity induced pairing
response $\langle D_{ij}\rangle $ is shown in  Fig.~\ref{fig:2}(b) for a
case with strong pairing.  The strength of
the induced pair field depends upon $J$, $t'$ and the doping $x$. 
As a measure of this response,
its magnitude  on the 12th $y=2-3$ rung
is plotted in Fig.~\ref{fig:2}(c) versus the hole
doping for various values of $t'$ and $J$. As previously found in
both DMRG and VMC  calculations, a positive value of $t'$ favors
pairing while a negative value  suppresses it.  Here one also
sees that when the pairing is strongest, the  response peaks
for a linear filling $\rho\sim0.5$ holes per unit length, but shifts
to higher doping for smaller $t'$. We see a strong dependence on $J/t$, with pairing when $J/t=0.3$
quite weak and when $J/t=0.5$ quite strong. Also,
as  shown in Fig.~\ref{fig:3}, the compressibility, which is related to the
slope of the curves, for $\rho=0.5$
increases as  $t'$ increases, consistent with the observed
enhancement of the pairing response  for positive values of $t'$.
The increased pairing for
larger $J/t$
may be due to a reduced repulsion between pairs, leading to an
enhanced compressibility.

\begin{figure}[thbp]
\begin{center}
\includegraphics[width=7cm,clip,angle=0]{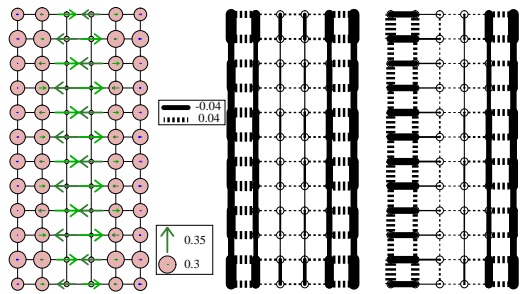}
\vskip 0cm
\caption{(a) Hole and
spin densities for a $12\times6$ ladder (plotted with $x$ oriented vertically) with $J=0.5$ and $t'=0$, and
with open boundary
conditions in both directions with fields applied to force two stripes.  The center
two legs have an applied staggered field $(-1)^x h$ with
$h=0.05$, along with a chemical potential of 1.3.  The outer four
legs have a chemical potential of 0.9, leading to a doping of
x=0.1981, corresponding to a linear doping per stripe of 0.594.
The outermost two legs have applied pair fields of 0.5 on each
horizontal rung. In (a), the pair fields are applied in phase.
(b) Measured pair fields for the in-phase case.
(c) Measured pair fields when the applied fields have opposite
sign.
}
\label{fig:4}
\end{center}
\end{figure}

\begin{figure}[thbp]
\begin{center}
\includegraphics[width=5cm,clip,angle=0]{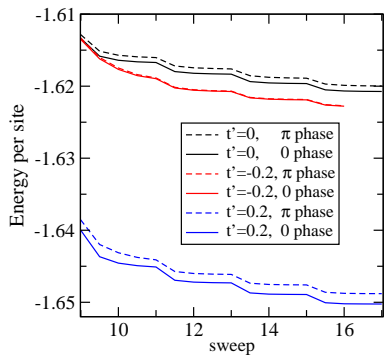}
\vskip 0cm
\caption{Energy per site for the $12\times6$ ladder of Fig. 4
versus sweep number for the in phase and antiphase applied
fields.  The runs all had $J=0.5$ and had the values of $t'$ indicated.
}
\label{fig:5}
\end{center}
\end{figure}

Next we turn to the question of anti-phase domain walls in the
pair field.  For values of $t'$ and doping where there is a
significant  pair field response, we find that the pair field
remains in phase across the stripes, as  shown in
Fig.~\ref{fig:1}(b). This implies that the energy to create an
anti-phase domain wall is positive. To probe this, we have
studied $12\times6$  ladders with open boundary conditions in
both the x and y directions, and with magnetic and chemical potential
fields applied to force two stripes. Fig.~\ref{fig:4}
shows such a system where each stripe has a linear doping of 
 $0.6$. Then by
applying pair fields on every link on the outermost legs ($y=1$ and $y=6$), a pair field is established. When the
applied  pair fields are in phase, the induced pair field is shown
in the middle figure of Fig.~\ref {fig:4}. The right hand figure shows
what happens when the applied pair fields are out of  phase. The
energy per unit length versus the DMRG sweep number for the in-phase
and anti-phase configurations are plotted in
Fig.~\ref{fig:5} for different values  of $t'$.  Although the energy difference is small,
it varies little with the sweep as the number of states in increased, up to $m=3000$ for sweep 17.
Here one sees
that it costs energy to create an anti-phase domain  wall and the
energy per unit length increases as the overall strength of the
induced pair field increases. For $t'=0.2$, the energy per unit length of the anti-phase domain wall is of order
0.01t.

There are both similarities  and differences between our DMRG results 
and those from VMC.\cite{ref:Himeda}  Both approaches find 
evidence for low lying striped
states with $d_{x^2- y^2}$ pair fields, as seen experimentally.
However, we find that negative values of
$t'$ suppress the $d$-wave  pair field.  Thus the parameter regime
where we have studied the interplay of  the stripes and the
pair field differs from the $t'<0$ region of the VMC study.
 Our DMRG calculations also find that  the
energy to form an anti-phase $d$-wave domain wall is positive.

In comparing to experiments for the cuprates, the puzzle raised by our
calculations,  as well as the RMFT\cite{ref:Yang} results, is the
positive energy required to  create an anti-phase domain wall in
the $d$-wave order. 
Because this is a small energy,
one can imagine that effects missing from the $t$-$t'$-$J$ model could lead to antiphase domain walls. 
However, we believe that there could be an alternate  explanation of the suppression of the interlayer
Josephson coupling observed in  the underdoped regime.
It could be that the decoupling arises from the lack of  overlap of the
Fermi surfaces of the adjacent layers. There are a number of
experiments\cite{ref:LeBoeuf,ref:Zabolotnyy} which imply that a
Fermi surface reconstruction occurs for hole doping near $x=1/8$.
The resulting  Fermi surface is characterized by
electron pockets and open orbits  whose location in the Brillouin
zone depends upon the stripe orientation.\cite {ref:Millis} In
this case, since the stripes alternate their orientation by $90^
\circ$ from one plane to the next, the lack of overlap between
the Fermi surfaces  can lead to a suppression of the interlayer
pair transfer processes\cite{ref:foot2}.

%

We wish to thank J.E.~Davis, S.A.~Kivelson and J.~Tranquada for  
insightful discussions. SRW acknowledges the support of the NSF under grant
DMR-0605444.


\begin{thebibliography}{99}

\bibitem{ref:Kohsaka} Y.~Kohsaka, C.~Taylor, K.~Fujita,  
A.~Schmidt, C.~Lupien, T.~Hanaguri, M.~Azuma, M.~Takano, H.~Eisaki, H.~Takagi,
S.~Uchida, J.C.~Davis, {\it Science} {\bf 315}, 1380 (2007)

\bibitem{ref:Tranqueda} J.M.~Tranqueda et al., {\it Nature} {\bf  
429}, 534 (2004).

\bibitem{ref:Abbamonte} P.~Abbamonte et al., {\it Nature} {\bf  
1}, 155 (2005).

\bibitem{ref:Kosterlitz} J.M.~Kosterlitz and D.J.~Thouless, {\it  
J. Phys. C} {\bf 6}, 1181 (1973).

\bibitem{ref:Li} Q.~Li, M.~Hucker, G.D.~Gu, A.M.~Tsvelik and  
J.M.~Tranqueda, {\it Phys. Rev. Lett.} {\bf 99}, 67001 (2007).

\bibitem{ref:Halperin} B.I.~Halperin and D.R.~Nelson, {\it J. Low  
Temp. Phys.} {\bf 36}, 599 (1979).

\bibitem{ref:Lake} B.~Lake, H.M.~Ronnow, N.B.~Christensen,  
G.~Aeppli, K.~Lefmann, D.F.~McMorrow,
P.~Vorderwisch, P.~Smeibidl, N.~Mangkorntong, T.~Sasagawa,  
M.~Nohara, H.~Takagi,
and T.E.~Mason, {\it Nature} {\bf 415}, 299 (2002).

\bibitem{ref:Schafgans} A.A.~Schafgans, private communication.

\bibitem{ref:Berg} E.~Berg, E.~Fradkin, E.-A.~Kim, S.A.~Kivelson,  
V.~Oganesyan, J.M.~Tranqueda
and S.C.~Zhang, {\it Phys. Rev. Lett.} {\bf 99}, 127003 (2007).

\bibitem{ref:Himeda} A.~Himeda, T.~Kato and M.~Ogata, {\it Phys.  
Rev. Lett.} {\bf 88}, 117001 (2002).

\bibitem{ref:Poilblanc} D.~Poilblanc and T.M.~Rice, {\it Phys.  
Rev. B} {\bf 39}, 9749 (1989);
H.J.~Schultz, {\it Phys. Rev. Lett.} {\bf 64}, 1445 (1990);
J.~Zaanen and O.~Gunnarsson, {\it Phys. Rev. B} {\bf 40}, 7391  
(1989).

\bibitem{ref:Machida} K.~Machida and M.~Ichioka, {\it J. Phys.  
Soc. Jpn.} {\bf 68}, 4020 (1999).

\bibitem{ref:WhiteScalapinoArxiv9610104} S.R.~White and  
D.J.~Scalapino, $d_{x^2-y^2}$
Pair Domain Walls, arXiv:cond-mat/9610104v1.

\bibitem{ref:WhiteScalapino3} S.R.~White and D.J.~Scalapino, {\it  
Phys. Rev. Lett.} {\bf 80}, 1272 (1998);
{\it Phys. Rev. B} {\bf 60}, R753 (1999); {\it Phys. Rev. Lett.}  
{\bf 81}, 3227 (1998).

\bibitem{ref:Yang} Kai-Yu Yang, Wei-Qiang Chen, T.M.~Rice,  
M.~Sigrist and Fu-Chun Zhang, arXiv:0807.3789v1

\bibitem{ref:Raczkowski} M.~Raczkowski, M.~Capello, D.~Poilblanc,  
R.~Fr\'esard and A.M.~Ole\'s,
{\it Phys. Rev. B} {\bf 76}, 140505(R) (2007).

\bibitem{ref:Capello} M.~Capello, M.~Raczkowski and D.~Poilblanc,  
{\it Phys. Rev. B} {\bf 77}, 224502 (2008).

\bibitem{ref:foot1} The experimental observation that the spin  
ordering temperature is coincident with the Josephson decoupling
of the CuO$_2$  planes suggests that
in fact the antiferromagnetic order plays an important role.

\bibitem{ref:Zabolotnyy} V.B.~Zabolotnyy, A.A.~Kordyuk, D.S.~Inosov,
D.V.~Evtushinsky, R.~Schuster, B.~Buechner, N.~Wizent, G.~Behr,  
Sunseng Pyon, H.~Takagi, R.~Follath and S.V.~Borisenko, arXiv:0809.2237v1

\bibitem{ref:LeBoeuf} D.~LeBoeuf \etal, {\it Nature} {\bf 450},  
533 (2007).

\bibitem{ref:Millis} A.J.~Millis and M.R.~Norman, {\it Phys. Rev.  B} 
{\bf 76}, 220503 (2007).

\bibitem{ref:foot2} A rough estimate of this suppression gives a  
factor of $(\Delta/t)^2$.

\end{thebibliography}
\end{document}